\documentclass[sigconf]{acmart}
\renewcommand\footnotetextcopyrightpermission[1]{} 
\usepackage[english]{babel}
\usepackage{blindtext}
\usepackage{multirow}

\renewcommand\footnotetextcopyrightpermission[1]{} 
\setcopyright{none}

\acmDOI{}

\acmISBN{}

\acmPrice{}

\begin{document}
\title{Integrated Access and Backhaul in 5G with Aerial Distributed Unit using OpenAirInterface}
\settopmatter{authorsperrow=1} 
\newcommand{\tsc}[1]{\textsuperscript{#1}} 
\author{Rakesh Mundlamuri\tsc{1}, Omid Esrafilian\tsc{1}, Rajeev Gangula\tsc{2}, Rohan Kharade\tsc{3}}
\author{Cedric Roux\tsc{1}, Florian Kaltenberger\tsc{1}, Raymond Knopp\tsc{1}, David Gesbert\tsc{1}}
\affiliation{
  \institution{\vskip .2cm}
  \institution{1. Communication Systems Department,
EURECOM, Biot, France}
  \institution{2. Institute for the Wireless Internet of Things, Northeastern University, Boston, USA}
  \institution{3. OpenAirInterface Software Alliance, Biot, France}
}
\renewcommand{\shortauthors}{Rakesh Mundlamuri.et al.}

\begin{abstract}
In this work, we propose an UAV-aided Integrated Access and Backhaul (IAB) system design offering 5G connectivity to ground users. UAV is integrated with a distributed unit (DU) acting as an aerial DU, which can provide 5G wireless backhaul access to a terrestrial central unit (CU). The CU-DU interface fully complies with the 3GPP defined F1 application protocol (F1AP). Such aerial DU can be instantiated and configured dynamically, tailoring to the network demands. The complete radio and access network solution is based on open-source software from OpenAirInterface (OAI) and off-the-shelf commercial 5G mobile terminals. Experimental results illustrate throughput gains and coverage extension brought by the aerial DU.


\end{abstract}

\maketitle

\section{Introduction}

Unmanned aerial vehicle (UAV) mounted base stations (BSs) and access points have recently attracted significant attention \cite{8660516}. Thanks to the 3D mobility offered by the UAV BSs, they are instrumental in providing ultra-flexible radio network deployments in use cases such as tactical networks, disaster recovery, search and rescue scenarios.

Several prototypes of UAV BSs and/or relays in 4G and 5G networks using open-source software are reported in the literature \cite{AuRelay,ChaAyonEug,9217782,10.1145/3495243.3558750}. While the UAV relay in \cite{ChaAyonEug,9217782} has an on-board core network and relies on commercial backhaul links, authors in \cite{AuRelay}, demonstrated an UAV LTE relay with integrated access and backhaul (IAB) capability. The work is later extended to 5G scenario \cite{mundlamuri}. However, in all these works entire eNB or gNB application is running on the UAV.

On the other hand, disaggregated radio access network (RAN) with open interfaces and end-to-end programmability has become an essential element in 5G and beyond cellular networks. Traditional gNB unit can now be split into various entities such as centralized unit (CU), distributed unit (DU) and radio unit (RU). CU contains packet data convergence protocol (PDCP) and above layers  in RAN protocol stack and can support multiple DUs running Radio link control layer and below network functionalities. With this architecture, DUs and RUs can be instantiated and programmed according to the needs of the network in a centralized manner \cite{polese}. 

In this paper, we present our system design of an UAV having the combined functionalities of DU and RU, serving ground users. The UAV is connected with a terrestrial CU using IAB. The CU-DU interface fully complies with the F1 application protocol (F1AP) defined in 3GPP. The end-to-end network solution is based on OpenAirInterface (OAI) software. 
To the best of our knowledge, the design and implementation of an aerial DU unit with IAB capabilities, built using open-source solutions, has never been demonstrated. 
\vspace{-3mm}
\section{System Design}
\begin{figure}[t]
\centerline{\includegraphics[width=3in]{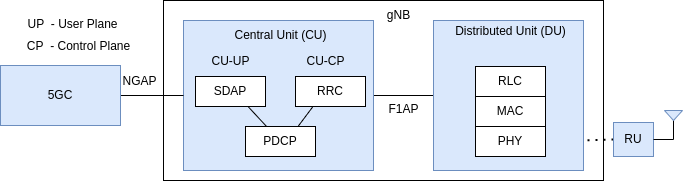}}\vspace{-3mm}
\caption{CU-DU split.}
\label{fig:graphic_cu_du_split}
\vspace{-4mm}
\end{figure}

\vspace{-1mm}
To support open and disaggregated radio access network (RAN), 3GPP TS 38.401 outlines the functional split of the gNB into central unit CU, DU and RU. The CU comprises radio resource control (RRC) and PDCP layer protocol stack in the control plane, and service data adaption protocol (SDAP) and PDCP in the user plane. The DU comprises RLC, medium access control (MAC), and/or physical (PHY) layers. RU consisting of PHY layer is either a part of DU itself or can have separate interface with DU. We consider the scenario where RU is a part of the DU. Each CU can be connected to one or more DUs. Communication between the CU and DU is carried out as a tunnel using F1AP as illustrated in Figure~\ref{fig:graphic_cu_du_split}. While the wired connection between CU and DU is straightforward, connecting the aerial DU with a terrestrial CU wirelessly over existing 5G RAN requires Integrated Access and Backhaul (IAB).

\begin{figure}[t]
\centerline{\includegraphics[width=3in]{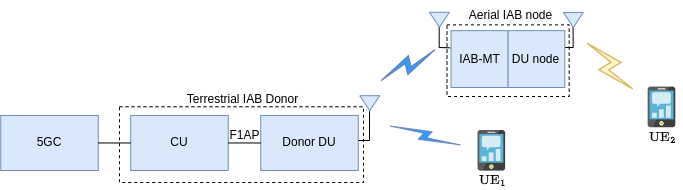}}\vspace{-3mm}
\caption{Block diagram of Integrated Access and Backhaul Scenario.}
\label{fig:graphic_Scenario}
\vspace{-7mm}
\end{figure}

An IAB system typically consists of an IAB Donor and IAB nodes \cite{9905510}. The IAB Donor, in our case, is a terrestrial CU wired to the DU referred to as Donor DU. The IAB node (which is mounted on a drone) consists of an IAB-Mobile Terminal (MT), and a DU. The DU within the IAB node is called as a DU node. The DU node can wirelessly connect to the CU in the IAB Donor via IAB-MT, as depicted in Figure~\ref{fig:graphic_Scenario}.

\begin{figure}[t]
\centerline{\includegraphics[width=2.9in]{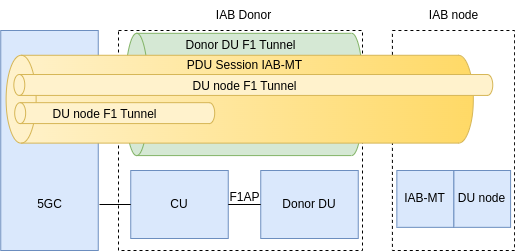}}\vspace{-2mm}
\caption{F1 tunneling in the setup.}
\label{fig:graphic_tunnel}
\vspace{-5mm}
\end{figure}

To achieve the over-the-air F1 interface, we implement F1 tunnel between the terrestrial CU and DU node as illustrated in Figure~\ref{fig:graphic_tunnel}.
This is achieved by using the General packet radio service Tunnelling Protocol (GTP) via IAB-MT. The F1 tunnel is created within the packet data unit (PDU) session of the IAB-MT. The F1 tunnel of the DU node starts via IAB-MT, passes over-the-air, tunnels through Donor DU, CU, and ends at 5GC User Plane Function (UPF). The packets are further rerouted from 5GC UPF to the CU, completing the F1 tunnel for the IAB node. However, the path to 5GC UPF from CU and the rerouting from the 5GC UPF to CU can be bypassed by utilizing the backhaul adaptation protocol (BAP) in the IAB node tunnel as mentioned in 3GPP TS 38.340. Although the BAP protocol has not been implemented in the OAI, the functionality of an IAB system can be demonstrated using the proposed solution. We plan to incorporate the BAP protocol in future implementations, ultimately improving the overall functionality of the system.
\vspace{-3mm}
\section{Setup description}

We present an IAB scenario in 5G with an aerial DU node serving ground users that can be deployed on-the-fly while being organized in a centralized manner. The proposed 5G IAB system comprises a Terrestrial IAB Donor and an Aerial IAB node that operate in different frequency bands. The Terrestrial IAB Donor is present on the ground and is made up of a CU and a Donor DU that are interconnected via Ethernet. 
The Aerial IAB node is a custom-built box that includes a commercial Quectel RM500Q-GL user equipment and a DU node mounted on an UAV. The DU node uses USRP B200 mini as an RU and a custom-built power amplifier to improve its coverage. Entire radio and access network solution is based on the OpenAirInterface (OAI) software. An illustration of the setup is shown in Figure~\ref{fig:graphic_Scenario}. The Terrestrial Donor DU uses 20MHz of bandwidth and operates at a center frequency of 2.585 GHz (n41), while the Aerial DU node uses 30MHz of bandwidth and operates at a center frequency of 3.47 GHz (n78). Both systems have a subcarrier spacing of 30 KHz.

Furthermore, a video recording\cite{mundlamuri_iab} has been made showcasing a live video call between users $\textrm{UE}_1$ and $\textrm{UE}_2$ as shown in Figure~\ref{fig:graphic_Scenario}. In this setup, $\textrm{UE}_2$ is out of coverage for direct link to the Donor DU, while by using the aerial DU, $\textrm{UE}_2$ can receive a throughput of 30 Mbps from 5GC.


\vspace{-3mm}
\section{Conclusion}
We have successfully implemented the 5G Aerial IAB scenario prototype using a complete open-source solution (OAI). The throughput results demonstrate the potential of the Aerial IAB system in extending coverage. Our prototype enables the researchers in the community to address a wide range of IAB-related research problems related to IAB and implement and verify them in real time. Future extensions include radio resource and interference management between Donor DU and aerial DU for in-band relaying, RAN Intelligent Controller (RIC)-based control, and joint optimization of aerial DU location, Bandwidth, and TDD configuration can be formulated based on the situational demand in O-RAN.
\vspace{-3mm}
\section{Acknowledgements}
This work was funded by 5G-OPERA, Imagine B-5G, and the German-French Academy for the Industry of the Future under project 3CSI.


\bibliographystyle{ACM-Reference-Format}
\bibliography{reference}

\end{document}